%% file: main.tex
\documentclass[10pt,conference]{IEEEtran} 

\usepackage{graphicx}
\usepackage{amsmath} 
\usepackage{amssymb}
\usepackage{comment} 
\usepackage{subcaption}
\usepackage{cite}
\usepackage[hidelinks]{hyperref}
\usepackage[percent]{overpic}

\title{Resource-Optimal Screening for Two-Tier Quantum System Characterization}

\author{
    Yi-Ting Chen\IEEEauthorrefmark{1}\IEEEauthorrefmark{3},
    Aniket S. Dalvi\IEEEauthorrefmark{2}\IEEEauthorrefmark{3},
    Nils Quetschlich\IEEEauthorrefmark{1}\IEEEauthorrefmark{3}
    \\
    \IEEEauthorblockA{\IEEEauthorrefmark{1}Amazon Braket, New York, NY, USA} 
    \IEEEauthorblockA{\IEEEauthorrefmark{2}Amazon Braket, Seattle, WA, USA} 
    \IEEEauthorblockA{\IEEEauthorrefmark{3}AWS Quantum Technologies, Seattle, WA, USA} 
}

\begin{document}

\maketitle

\begin{abstract}
Quantum processing units (QPUs) require frequent benchmarking to track hardware drift, yet standard protocols such as randomized benchmarking are resource-intensive and compete with application workloads for scarce QPU time. We present a resource-optimal screening strategy that combines a two-tier, risk-based testing architecture with optimal experiment design: a lightweight screening probe runs regularly and triggers expensive full characterization only when statistically significant change is detected, while the probe itself is designed to maximize detection sensitivity under explicit resource constraints. For an exponential fidelity model, we derive closed-form optimal circuit sizes and introduce an effective gate count that renders the analysis hardware-independent. We instantiate the strategy for two applications. For gate-quality verification, we show that optimally designed screening probes outperform randomized benchmarking with optimal experimental design (OED) by more than 4$\times$ in detection rate at equal resource cost. For compiler benchmarking, we formalize compiler comparison as a paired hypothesis test and derive the optimal circuit size for distinguishing compilation strategies. Despite their different observables and test statistics, both applications yield structurally the same closed-form expression for the optimal probe circuit size as a function of gate fidelity, illustrating the generality of the framework. Numerical and hardware experiments on QPUs (Rigetti Ankaa-3, IQM Garnet, IQM Emerald) confirm the theoretical predictions, with observed optimal circuit sizes matching the derived values across all devices.
\end{abstract}


\section{Introduction}

Quantum processors are subject to continuous parameter drift: control fields, qubit coherence times, and noise spectra change on timescales of hours to days~\cite{Proctor2020,Klimov2018}. Without recalibration, error rates can degrade rapidly; for example, on a superconducting device, over 90\% of single-qubit gates exceeded the ${\sim}1\%$ error threshold needed for surface-code error correction after just one day~\cite{fang2025caliqec}. Hardware vendors mitigate this through periodic calibration routines, typically daily and lasting 30--90 minutes per system~\cite{ibm2024calibration_jobs}, with some providers calibrating only weekly~\cite{zhu2024leveraging}.

Benchmarking must therefore be repeated frequently to track drift, yet the tasks that constitute a complete benchmarking workflow are diverse and resource-intensive. Gate-quality characterization via randomized benchmarking (RB) requires ${\sim}10^{2}$ random sequences at multiple lengths with ${\sim}10^{3}$ shots each~\cite{knill2008rb,magesan2012characterizing}. Error mitigation techniques such as probabilistic error cancellation~\cite{temme2017error,vandenberg2023pec} and zero-noise extrapolation~\cite{li2017efficient,he2020} require accurate noise models whose learning cost scales with circuit size~\cite{mcdonough2022,piveteau2022,kim2023evidence}. Compiler benchmarking, which compares how different compilation strategies affect execution fidelity on a given device, demands QPU runs of representative workloads~\cite{lubinski2023benchmarks,tomesh2022supermarq}. Quantum error correction (QEC) decoders that exploit device error models achieve lower logical error rates for fault-tolerant applications~\cite{chen2022calibrated,battistel2023real}. Noise-adaptive compilers themselves depend on current calibration data~\cite{murali2019noiseadaptive,tannu2019noiseadaptive}. The resource pressure is compounded in the cloud-access model, where benchmarking competes with application workloads for queue time.

Despite their diversity, benchmarking tasks share a common statistical structure: each can be modeled as an estimation of a quality parameter (a gate fidelity, a noise-model coefficient, a compiler-induced fidelity difference) from finite measurement data, subject to resource constraints. Whether one is monitoring component-level gate fidelities~\cite{knill2008rb,magesan2011scalable}, system-level algorithm success rates~\cite{emerson2005scalable,cross2019validating,kandala2017hardware,cowtan2020tket}, or compiler effectiveness~\cite{cross2019validating}, the central challenge is the same: maximizing the information gained per unit of QPU time.

\begin{figure*}[t]
    \centering
    \begin{overpic}[width=0.9\linewidth]{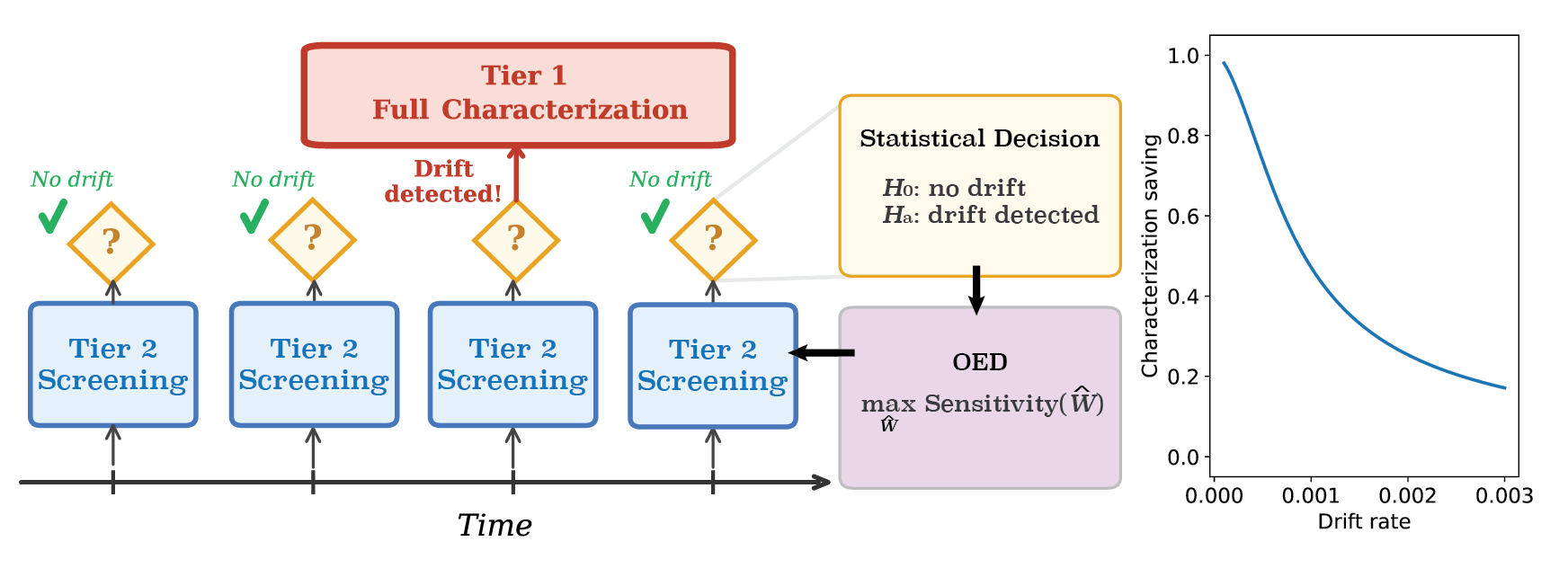}
        \put(2,33){\textbf{(a)}}
        \put(71,33){\textbf{(b)}}
    \end{overpic}
    \caption{Overview of the two-tier characterization strategy. (a) Two-tier screening architecture. A lightweight OED-optimized Tier-2 screening probe runs at every benchmarking interval. Only when statistically significant drift is detected does it trigger the expensive Tier-1 full characterization, with screening cost much lower than full characterization. (b) Saving $\eta(\sigma)$ of the sense-and-characterize strategy relative to always-characterize, as a function of drift rate $\sigma$.}
    \label{fig:risk_based_testing}
\end{figure*}

In this work, we develop a resource-optimal screening strategy that exploits this shared structure (Fig.~\ref{fig:risk_based_testing}a). The strategy combines \emph{risk-based testing}, replacing routine full characterization with a lightweight screening probe that triggers the expensive procedure only upon detecting statistically significant change, with \emph{optimal experiment design} (OED) to maximize the probe's sensitivity under explicit resource constraints. Our main contributions are:
\begin{enumerate}
    \item We derive closed-form optimal probe circuit sizes under an exponential fidelity model by maximizing detection sensitivity, and connect the result to Fisher information maximization. We further introduce the effective gate count (EGC) to make the analysis hardware-independent (Section~\ref{sec:screening_strategy}).
    \item We instantiate the framework for two applications, quality verification (Section~\ref{sec:app_quality}) and compiler benchmarking (Section~\ref{sec:compiler_benchmarking}), and show that, despite their different observables and test statistics, both yield structurally the same closed-form expression for the optimal circuit size as a function of gate fidelity. For quality verification, the optimally designed screening probe outperforms OED-optimized RB by more than $4\times$ in detection rate at equal resource cost. For compiler benchmarking, we formalize compiler comparison as a paired hypothesis test on Hellinger fidelity differences and derive the optimal circuit size for distinguishing compilation strategies.
    \item We validate the theoretical predictions with hardware experiments targeting QPUs (Rigetti Ankaa-3, IQM Garnet, IQM Emerald) and numerical simulations on IonQ Aria-1, confirming that observed optimal circuit sizes match the derived values across all devices.
\end{enumerate}

\section{Background}

\subsection{Risk-based testing}

Risk-based testing is a two-step methodology: a lightweight screen identifies the highest-risk components, and expensive characterization is directed only toward flagged items. This paradigm has been formalized in software engineering~\cite{felderer2014taxonomy} and adopted across industries including semiconductor test, drug discovery, aerospace, and manufacturing, where cheap screens (parametric measurements, virtual assays, low-fidelity simulations, sensor-stream anomaly detection) gate expensive procedures~\cite{Soden1993iddq,sumikawa2011forward,gorgulla2020virtualflow,peherstorfer2018survey,toal2015multifidelity,canizo2019anomaly,saxena2008turbofan}, yielding cost reductions of up to 29-fold in drug discovery~\cite{elez2025simulations}.

Quantum system benchmarking faces an analogous challenge: full RB characterization~\cite{knill2008rb,magesan2012scalable,helsen2022general}, noise-model learning for error mitigation~\cite{vandenberg2023pec}, and fidelity-based compiler evaluation~\cite{lubinski2023benchmarks,tomesh2022supermarq} are all resource-intensive. The closest quantum-specific precedent is the Optimus algorithm of Kelly et al.~\cite{kelly2018physical}, which organizes qubit calibrations as a dependency graph and gates full calibration behind lightweight checks; Riesebos et al.~\cite{riesebos2021universal} integrate this architecture into a production scheduling toolkit. In both works, however, the screening tier is heuristic, with no prescription for how the screening probe itself should be designed. 

\subsection{Optimal experiment design for quantum benchmarking}

Optimal experiment design (OED) seeks to maximize the information extracted from experiments under a limited budget of measurements or circuit executions. In the context of RB, Kwiatkowski et al.~\cite{kwiatkowski2023} and Itoko and Raymond~\cite{itoko2021sampling} optimize sequence-length and sample allocations to minimize estimator variance, Harper et al.~\cite{harper2019} quantify how design choices impact error-rate precision, and Hincks et al.~\cite{hincks2018} develop a Bayesian framework for principled data-collection decisions. Hashagen et al.~\cite{hashagen2018} extract richer structural information at comparable cost via real RB, and Kelly et al.~\cite{kelly2014} use RB-derived fidelity as a closed-loop calibration signal in the ORBIT protocol.

For compiler benchmarking, most existing work provides benchmark suites such as Arline Benchmarks~\cite{kharkov2022}, SupermarQ~\cite{tomesh2022supermarq}, and MQT Bench~\cite{quetschlich2023}, but does not optimize the evaluation itself for maximum sensitivity under resource constraints. Since compilation quality directly affects execution fidelity~\cite{botea2018,siraichi2019,li2019,sivarajah2020}, and production compilers rely on heuristics whose relative performance varies across devices, the application of OED to compiler benchmarking remains a largely open direction.

\section{Optimal Screening Design}
\label{sec:screening_strategy}

We now present the optimal screening strategy, which is task-agnostic: it applies to any benchmarking problem that can be cast as a statistical decision problem with a tunable probe circuit. We describe the two-tier architecture, formalize the decision problem, derive the optimal probe design, introduce the hardware-independent effective gate count, and state the constrained optimization a practitioner solves in practice.

\subsection{Two-tier screening strategy}
\label{sec:two_tier}

We structure quantum system screening as a two-tier process:
\begin{itemize}
    \item \textbf{Tier~1 (full characterization):} The expensive full benchmarking procedure (e.g., complete RB, noise-model learning, or compiler investigation) is triggered only when the Tier~2 screen raises an alarm.
    \item \textbf{Tier~2 (screening):} A lightweight probe measurement is performed at regular intervals. The probe has cost $C_{\mathrm{probe}} \ll C_{\mathrm{full}}$ and produces a test statistic that is compared against a threshold.
\end{itemize}

To demonstrate the effectiveness of the two-tier process, we abstractly model a quality parameter $\theta_i$ as undergoing a discrete random walk $\theta_i = \theta_{i-1} + x_i$, $x_i \sim \mathcal{N}(0, \sigma^2)$, starting from $\theta_0 = 0$ (freshly calibrated). The Tier~2 sensor raises an alarm whenever $|\theta_i| > \delta$, where $\delta$ is the detection threshold. Upon alarm, full characterization resets $\theta_i = 0$.

The saving relative to an always-characterize baseline is $\eta(\sigma) = 1 - \langle N_{\mathrm{cal}}(\sigma) \rangle / N_{\mathrm{steps}}$, where $\langle N_{\mathrm{cal}} \rangle$ is the expected number of Tier~1 triggers over $N_{\mathrm{steps}}$ screening intervals. Fig.~\ref{fig:risk_based_testing}b shows $\eta(\sigma)$ estimated via Monte Carlo simulation. When the drift rate is small relative to the threshold ($\sigma \ll \delta$), the regime most relevant to current quantum processors, nearly all full characterizations are avoided and $\eta \to 1$. As $\sigma$ increases past $\delta$, the sensor triggers at every step and $\eta \to 0$, recovering the always-characterize baseline. 
The two error modes of the screen have asymmetric consequences: a false positive merely triggers an unnecessary Tier~1 run, whereas a false negative leaves a drifted device operating under an outdated calibration, the operationally dangerous outcome. Controlling the false-negative rate is therefore integral to the probe design. The detection power of the screen grows with the probe's sensitivity, so for a fixed budget, the sensitivity-maximizing probe derived below is also the one with the lowest false-negative rate. Operational safeguards that further bound this risk are discussed in Section~\ref{sec:discussion}.
Because the total resource expenditure is dominated by the Tier~2 screening cost in the low-drift regime, the central design question becomes: \emph{how should the screening probe be designed to maximize its sensitivity to drift?} The remainder of this section answers this question.

\subsection{General statistical decision problem}
\label{sec:general_hypothesis}

Every Tier~2 screening probe has the same abstract structure. An experimenter executes a probe circuit with a controllable design parameter $M$, the circuit size in effective gates (Section~\ref{sec:universality}), and observes an outcome $Y$ that depends on $M$ and an unknown quality parameter $\theta$. A reference value $Y_{\mathrm{spec}}$ is available from the device specification or a previous full characterization. The screening question, \emph{has the system quality $\theta$ changed since the last characterization?}, is formalized as a two-sided hypothesis test of $H_0 : \mathbb{E}[Y(M, \theta)] = Y_{\mathrm{spec}}$ against $H_a : \mathbb{E}[Y(M, \theta)] \neq Y_{\mathrm{spec}}$.
The decision rule compares a $t$-statistic $t = (Y - Y_{\mathrm{spec}})/\sigma_Y$ against a critical value $t_\alpha$, where $\sigma_Y$ is the standard error of the measurement and $\alpha$ is the chosen significance level.

This template encompasses diverse benchmarking tasks. Table~\ref{tab:instantiations} shows how quality verification and compiler benchmarking each instantiate the general formulation.

\begin{table}[h]
\centering
\caption{Instantiation of the screening strategy for two benchmarking tasks.}
\label{tab:instantiations}
\begin{tabular}{lcc}
\hline
 & \textbf{Quality verification} & \textbf{Compiler benchmarking} \\
\hline
Observable $Y$ & Success probability & Paired fidelity difference \\
Design param.\ $M$ & Clifford seq.\ length $N_g$ & Circuit size \\
Quality param.\ $\theta$ & Gate fidelity $\lambda$ & Compiler quality gap \\
$Y_{\mathrm{spec}}$ & $Y(\lambda_{\mathrm{spec}}, M)$ & $0$ (no difference) \\
$H_0$ & No drift & Compilers equivalent \\
\hline
\end{tabular}
\end{table}

\subsection{Optimal probe design via sensitivity maximization}
\label{sec:optimal_probe}

Under $H_0$, the expected $t$-value is zero regardless of the design parameters. Under $H_a$, however, the experimenter's choice of design parameters directly controls the sensitivity of the test. The goal is therefore to choose the design parameters to maximize $|t|$ under $H_a$.
In general, the design may involve multiple parameters (gate count, number of circuits, shots per circuit, etc.); the constrained optimization over the full parameter vector is treated in Section~\ref{sec:constrained_opt}. Here we focus on the circuit-size parameter $M$, which governs the probe's sensitivity and exhibits the most interesting structure.

For a general exponential fidelity model, widely used to describe circuit fidelity under depolarizing noise~\cite{knill2008rb,magesan2012characterizing},
\begin{equation}
    \label{eqn:fidelity_model}
    \mathbb{E}[Y] = (1-c)\,f_0^{M} + c,
\end{equation}
where $f_0 \in (0,1)$ is the base-gate fidelity, $M$ is the effective circuit size, and $c \in [0,1)$ is a floor parameter (e.g., the depolarization limit), a small drift $\Delta\theta$ in the quality parameter $\theta = f_0$ produces a signal $\mathbb{E}[Y] - Y_{\mathrm{spec}} \approx (\partial Y / \partial \theta)\,\Delta\theta = (1-c)\,M f_0^{M-1} \Delta\theta$. A qualitative optimum emerges from two competing effects: at small $M$ the probe is insensitive, since errors have not accumulated enough for the signal to exceed shot noise, while at large $M$ the expected fidelity $f_0^M$ decays exponentially, washing out the signal and inflating the relative variance.
The optimal $M$ balances these effects by maximizing the sensitivity-to-noise ratio $|\partial Y / \partial \theta| / \sigma_Y$.
This objective is equivalent to maximizing the Fisher information $\mathcal{I}(\theta; M)$ about $\theta$ at the design point $M$~\cite{Kay1998,Pukelsheim2006,VanTreesBellTian2013}. By the Cram\'{e}r--Rao bound, no unbiased estimator achieves variance below $\mathcal{I}^{-1}$, so maximizing $\mathcal{I}$ directly minimizes the fundamental limit on estimation error, connecting our strategy to classical optimal experiment design theory. The Fisher information analysis for both peaked distributions (quality verification) and broad distributions (compiler benchmarking) is summarized in Appendix~\ref{app:fisher}.

\subsection{Universality via effective gate count}
\label{sec:universality}

The analysis above assumes a single gate type with fidelity $f_0$. Real devices have heterogeneous error rates: every qubit and every coupler edge can have a distinct fidelity, and a circuit may contain an arbitrary mix of gate types. Under the assumptions of time-invariant and depolarizing noise, all such circuits can be mapped to an equivalent single-parameter model through the \emph{effective gate count} (EGC). The EGC is equivalent to a log-rescaling of the estimated success probability that has been widely used as a figure of merit in noise-aware compilation~\cite{tannu2019noiseadaptive,nishio2020extracting}. It is also related to the error per layered gate (EPLG) metric~\cite{mckay2023benchmarking}, but whereas EPLG is a top-down measure extracted from whole-circuit decay rates, the EGC is built bottom-up from individual gate fidelities. We adopt this aggregation because it linearizes the exponential fidelity model and makes the OED analysis hardware-independent.

Consider a circuit composed of $K$ gates, where the $k$-th gate has fidelity $f_k$ (which may differ for every qubit, edge, and gate type). The effective gate count relative to a reference fidelity $f_0$ is the sum of individual gate contributions:
\begin{equation}
    \label{eqn:egc}
    N_{\mathrm{EGC}} = \sum_{k=1}^{K} \frac{\log f_k}{\log f_0}.
\end{equation}
Each term $\log f_k / \log f_0$ converts the $k$-th gate's error into an equivalent number of reference gates, so the EGC aggregates the total noise burden of the circuit regardless of the heterogeneity of the underlying error rates. The expected fidelity then collapses to $F = f_0^{N_{\mathrm{EGC}}}$, reducing the multi-parameter model to a single-parameter exponential decay.

This abstraction makes the strategy universal. The optimal design analysis is performed once in the abstract space of effective gates; the resulting $M^\star_{\mathrm{EGC}}$ depends only on $f_0$ and is independent of the particular mix of gate types or qubit-level error variation. To apply the result to a specific device, one inverts the EGC mapping: given the per-gate fidelities and the circuit structure, $M^\star_{\mathrm{EGC}}$ translates back into a target number of physical gates. As device characteristics change or new gate types are introduced, only the translation needs updating; the optimality analysis remains unchanged.

\subsection{Constrained optimization}
\label{sec:constrained_opt}

Section~\ref{sec:optimal_probe} identified the optimal circuit size $M$ by maximizing detection sensitivity as a single-variable problem. In practice, $M$ is only one of several experimental parameters: the number of shots $N_s$ and circuits $N_c$ also affect detection sensitivity through the standard error $\sigma_Y$, and all parameters consume resources. The constrained optimization below jointly optimizes over all parameters under a resource budget.

We denote the vector of experimental parameters as $\hat{W}$ (encompassing the design parameter $M$, shots per circuit $N_s$, and number of circuits $N_c$) and assume that cost models for time and monetary expenditure, $\mathcal{T}(\hat{W})$ and $\mathcal{M}(\hat{W})$, are given. The operational output of the strategy is the constrained optimization problem
\begin{align}
    \label{eqn:general_optimization}
    &\max_{\hat{W}} \;\; \Phi(\hat{W}; \theta_{\mathrm{spec}}) \notag\\
    &\text{subject to} \quad \mathcal{T}(\hat{W}) \le T_{\mathrm{bound}}, \notag\\
    &\phantom{\text{subject to}} \quad \mathcal{M}(\hat{W}) \le M_{\mathrm{bound}},
\end{align}
where $\Phi$ is the task-specific sensitivity objective (the $t$-value for quality verification, or the expected fidelity difference for compiler benchmarking), $\theta_{\mathrm{spec}}$ denotes the device parameters from the specification, and $T_{\mathrm{bound}}$, $M_{\mathrm{bound}}$ are the time and monetary budgets. This formulation makes explicit what the na\"ive ``just run some circuits'' approach leaves implicit: every experimental choice has a cost, and the goal is to extract the maximum statistical information per unit of resource.

\section{Application 1: Spec verification}
\label{sec:app_quality}
In this section, we apply the strategy towards quality verification, detecting whether a QPU's gate fidelities have drifted from their published specification.

\subsection{Hypothesis test}
\label{sec:quality_hyp_test}

A hardware provider publishes a specification $\theta_{\mathrm{spec}}$ characterizing gate fidelities at calibration time. The screening question is: \emph{does the current QPU still agree with the specification?} Running a full RB experiment to re-estimate $\theta$ is wasteful when the goal is merely to detect drift. We instead formalize the problem as a hypothesis test following the template of Section~\ref{sec:general_hypothesis}:
\begin{align}
    H_0 &: \bar{Y} = S_{\mathrm{spec}}, \label{eqn:t-test quality verification}\\
    H_a &: \bar{Y} \neq S_{\mathrm{spec}},
\end{align}
where $\bar{Y}$ is the measured success probability of a probe circuit and $S_{\mathrm{spec}}=S(\theta_{\mathrm{spec}})$ is the success probability predicted by the specification. Rejection of $H_0$ is a necessary condition for $\theta_{\mathrm{QPU}} \neq \theta_{\mathrm{spec}}$: if the underlying fidelities have not changed, the measured success probability must be consistent with the specification. Only when $H_0$ is rejected does one invest in a full RB re-characterization.

\subsection{Design of the screening probe}
\subsubsection{RB protocol (Tier~1)}
Gate fidelities published by hardware providers are typically characterized by randomized benchmarking (RB)~\cite{knill2008rb,magesan2012characterizing}. We briefly review the RB protocol to establish notation and then describe how the screening probe differs from it. In an RB experiment, each circuit $i$ consists of a randomly sampled sequence of $N_g$ distinct Clifford gates $\{U_{i,1},U_{i,2},\ldots,U_{i,N_g}\}$, followed by an inversion gate that maps the ideal output back to the zero state. Under the standard assumptions of time-independent, gate-independent, Markovian noise, the noisy circuit is $\tilde{\mathcal{C}}_i = (\Lambda\circ U_{i,1})\circ (\Lambda\circ U_{i,2})\circ \cdots \circ (\Lambda\circ U_{i,N_g})$.
The noise channel $\Lambda$ can be modeled by a depolarizing channel with a single parameter $\lambda$~\cite{magesan2011scalable}. The expected success probability of a sequence of length $N_g$ is
\begin{equation} \label{eqn:noise_model}
S(\lambda, a, N_g)=a (1-c) \lambda^{N_g} + c
\end{equation}
where $a$ is the SPAM fidelity, $c$ is a constant depending on the qubit count, and $\lambda$ is the depolarizing parameter (the concrete instantiation of the general quality parameter $\theta$). The metric is the success probability of measuring the initial state, typically prepared as the zero state $|0\rangle^{\otimes n}$, at the output of a Clifford sequence circuit. This quantity corresponds to the Hellinger fidelity between the ideal and measured output distributions. 

\subsubsection{Probe circuits (Tier~2)}
\label{sec:probe_circuits}
Like RB, each probe circuit consists of a randomly sampled Clifford gate sequence of length $N_g$ followed by an inversion gate. The key difference is in how the protocol is used: RB samples multiple circuits at each of several sequence lengths to fit the exponential decay and estimate $\lambda$, requiring a large total number of circuits. The probe, by contrast, fixes $N_g$ at the single optimal value $M^\star$ and asks only whether the observed success probability is consistent with the specification. Multiple probe circuits, each with an independently sampled Clifford sequence of the same length $N_g$, are run to reduce statistical uncertainty, but because only one depth is needed the total circuit count is much smaller than in full RB. The probe's goal is \emph{detection} rather than \emph{estimation}. Under the same depolarizing noise model, the expected success probability of each probe circuit follows Eq.~\eqref{eqn:noise_model}.

Note that the Clifford sequence length $N_g$ counts the logical Clifford gates in a probe sequence, each of which decomposes into native physical gates whose fidelities determine its EGC contribution via Eq.~\eqref{eqn:egc}. For the analysis that follows, we choose the EGC reference fidelity to be the per-Clifford depolarizing parameter, $f_0 = \lambda$, so that each Clifford gate contributes exactly one effective gate and $M = N_g$. This is a choice of units rather than an approximation, and it lets us carry out the optimization directly in terms of $N_g$. When reporting device-specific gate counts against a different reference fidelity, we convert through Eq.~\eqref{eqn:egc}.

\subsection{Optimal experiment design} \label{sec:quality verification, optimal exp design}
\subsubsection{Decision rule}
After collecting data, we compute the $t$-statistic
\begin{equation}
    \label{eqn:t_statistic_decision_rule}
    t = \frac{\bar{Y} - S_{\mathrm{spec}}}{\sigma},
\end{equation}
where $\sigma$ is the standard error of the measurement.
Because the success probability is a binomial proportion estimated from $N_s$ shots per circuit and averaged over $N_c$ circuits, the standard error is $\sigma = \sqrt{{S(1-S)}/{N_s\,N_c}}$.
Given a significance level $\alpha$ with corresponding critical value $t_\alpha$, the decision rule is: reject $H_0$ if $|t| > t_\alpha$.

Following Section~\ref{sec:optimal_probe}, we seek the experimental parameters $(N_g, N_s, N_c)$ that maximize $|t|$ under $H_a$, subject to the resource constraints of Eq.~\eqref{eqn:general_optimization}.

\begin{figure}
\centering
\includegraphics[width=0.8\linewidth]{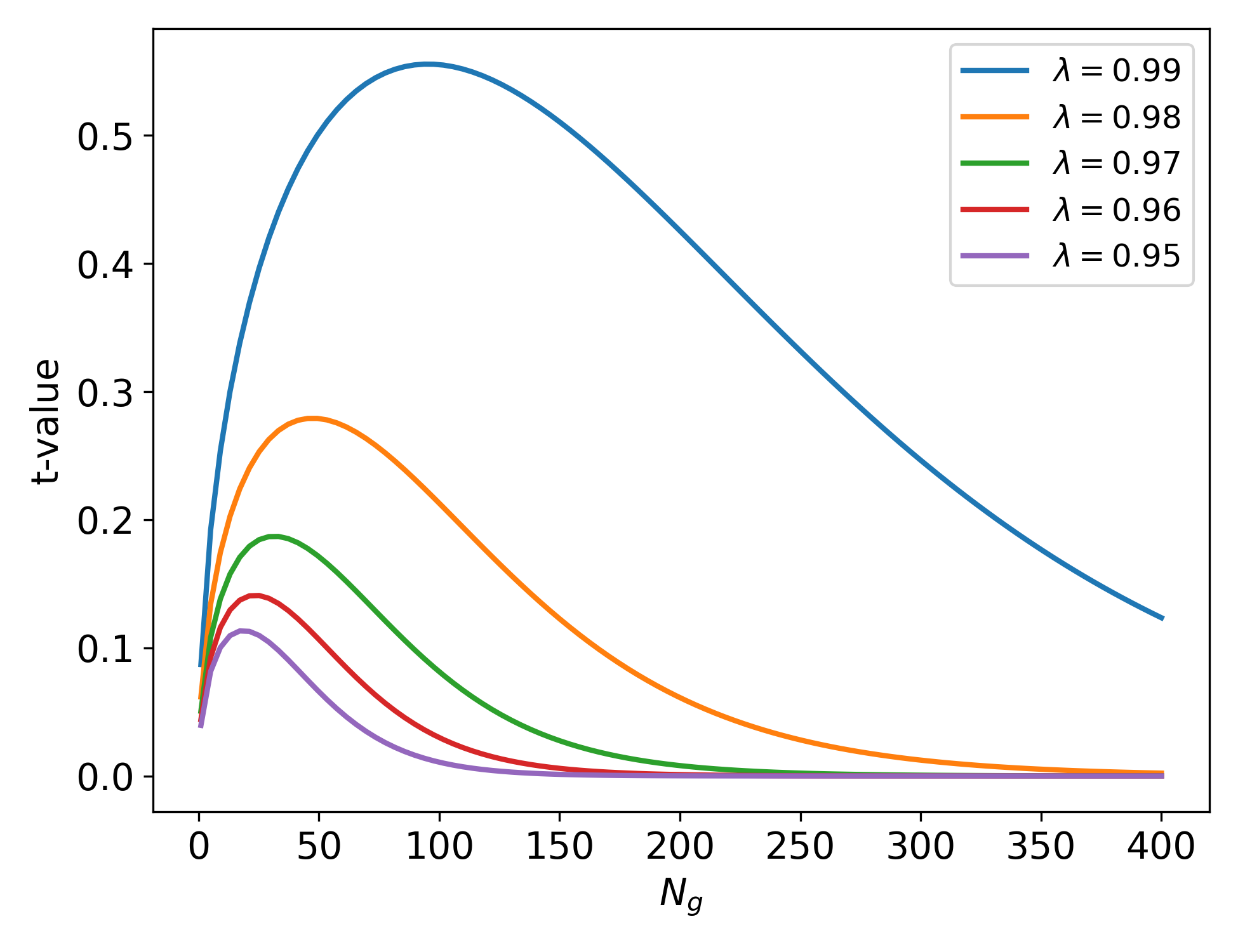}
\caption{Expected $t$-value of the screening probe (Eq.~\eqref{eqn:t_value_2}) as a function of Clifford sequence length $N_g$, for base fidelities $\lambda$ from $0.95$ to $0.99$. Peak positions match the closed-form optimum of Eq.~\eqref{eqn:optimal_M_general}. Parameters: $N_s=100$ shots, $N_c=1$ circuit, drift $\Delta\lambda=10^{-3}$ (representative of daily drift on current hardware), SPAM fidelity $a=1$ (to isolate the effect of $\lambda$), and depolarization floor $c=0.25$ for the two-qubit case.}
\label{fig:optimal_t_value}
\end{figure}

We model hardware drift after a time interval $\Delta T$ as small perturbations to the calibrated parameters:
$\lambda_{\mathrm{QC}} = \lambda_{\mathrm{spec}} + \Delta\lambda$ and
$a_{\mathrm{QC}} = a_{\mathrm{spec}} + \Delta a$,
with $\Delta\lambda \ll \lambda_{\mathrm{spec}}$ and $\Delta a \ll a_{\mathrm{spec}}$.
Expanding the $t$-value to first order in the drifts shows that the contribution of $\Delta a$ is suppressed by a factor $1/N_g$ relative to that of $\Delta\lambda$ and can be neglected (Appendix~\ref{app:t_value_expansion}), leaving
\begin{equation} \label{eqn:t_value_2}
    t = (1-c)\frac{
        N_g a \lambda^{N_g-1}\Delta\lambda
    }{\sigma}
    \Big|_{(\lambda_\text{spec}, a_\text{spec})}.
\end{equation}
Fig.~\ref{fig:optimal_t_value} plots the $t$-value of Eq.~\eqref{eqn:t_value_2} as a function of the Clifford sequence length $N_g$, with one curve per base fidelity $\lambda \in [0.95, 0.99]$. Every curve exhibits the rise-and-fall shape anticipated in Section~\ref{sec:optimal_probe}: the signal $N_g\,\lambda^{N_g-1}\Delta\lambda$ initially grows because each additional gate compounds the drift, then collapses once the exponential decay $\lambda^{N_g}$ drives the success probability toward the depolarization floor, where the drift signal is buried in shot noise. The competition between the linear amplification factor $N_g$ and the exponential attenuation $\lambda^{N_g-1}$ produces a unique interior maximum. Lower-fidelity devices reach this turnover sooner: the peak shifts from $N_g = 95$ at $\lambda = 0.99$ down to $N_g = 19$ at $\lambda = 0.95$, in agreement with the Fisher information result that the optimal gate count is
\begin{equation}
    \label{eqn:optimal_M_general}
    M^\star = \frac{x^\star(c)}{-\ln f_0} ,
\end{equation}
where $x^\star(c)\approx 0.96$ when $c=0.25$ (Appendix~\ref{app:fisher}). The figure thus provides a direct visual confirmation of the closed form: the peak locations of the numerically evaluated $t$-value match $M^\star$ across the full fidelity range, which is the optimality claim the figure is meant to support.

\subsubsection{Maximization under constraints}
In Eq.~\eqref{eqn:t_value_2}, $\Delta\lambda$ is unknown and it is not a tunable parameter. Therefore, when maximizing the t-value, we ignore $\Delta\lambda$ and denote the sensitivity factor as $t_s$:
\begin{equation} \label{eqn:quality verification, differential t value}
    t_s = (1-c)\frac{
        N_g a \lambda^{N_g-1}
    }{\sigma}
    \Big|_{(\lambda_\text{spec}, a_\text{spec})}
\end{equation}
This is the task-specific objective $\Phi$ in the general constrained optimization of Eq.~\eqref{eqn:general_optimization}, with $\hat{W} = (N_g, N_s, N_c)$.

\begin{figure*}[t]
\centering
\begin{overpic}[width=0.9\linewidth]{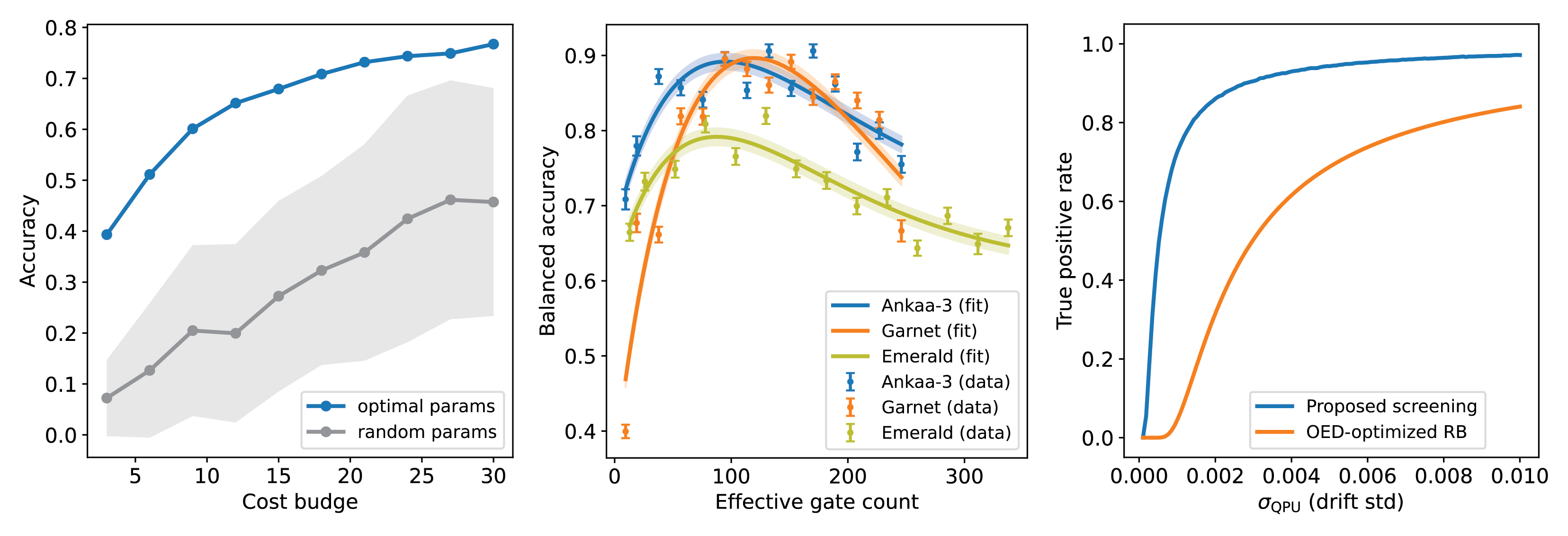}
    \put(0,33){\textbf{(a)}}
    \put(34,33){\textbf{(b)}}
    \put(66.5,33){\textbf{(c)}}
\end{overpic}
\caption{Detection performance of the optimally designed screening probe. (a) Accuracy of the optimal parameters and random parameters in numerical simulation. The experiment assumes $\lambda_\text{QC}=0.98$ and $a_\text{QC}=0.99$. The evaluation of the accuracy assumes the spec is a normal distribution with mean $0.98$ and std $10^{-3}$. The critical value is $t_\alpha=2$. The cost budget is set to both time and monetary cost. The gray area around the gray curve is the std over 40 sets of random parameters. (b) Balanced accuracy of the screening probe on QPUs as a function of effective gate count. Each data point is the average over 10 Clifford sequence experiments, each run with 2000 shots. (c) True positive detection rate as a function of drift rate $\sigma$ for the OED-optimized RB baseline~\cite{itoko2021sampling} and the proposed screening-probe approach. Both use equal resource budgets (68 circuits).}
\label{fig:quality_bench_results}
\end{figure*}

\subsection{Evaluate with numerical simulations}

In this section, we validate how effectively the optimally designed screening probe detects device drift. We use the true positive rate (TPR, ratio of true positives to actual positives) as the metric, an empirical measure of the screen's detection power (Section~\ref{sec:two_tier}). To isolate the benefit of optimal experiment design, we compare the optimally designed screening probe against a baseline in which the screening probe parameters $(N_g, N_s, N_c)$ are chosen uniformly at random, subject to the same resource bound.

We assume a hardware-inspired readout fidelity of $a_\text{QPU}=0.99$ and a drifting Clifford gate fidelity, $\lambda_\text{QPU}$, with mean $0.98$ and std $\sigma_\text{QPU}=10^{-3}$. 
We also assume that the drift of hardware follows a normal distribution. That is, after a time interval, $\lambda_\text{QPU}$ drifts from $\lambda_\text{spec}$ to $\lambda_\text{spec}+\Delta\lambda$ and $\Delta\lambda$ follows $\Delta\lambda \sim \mathcal{N}(0, \sigma_\text{QPU}^2)$ where $\mathcal{N}$ is the normal distribution.
The monetary cost model of the device is $\mathcal{M}(N_s, N_c)=0.3N_c+0.0009N_s$
where the unit is \$. 
The time cost is $\mathcal{T}(N_s, N_c)=0.003N_c+0.0002N_s$ where the unit is second. 

We sample many $\Delta\lambda$ and compute TPR, which we denote as the accuracy of the numerical experiment. The result is shown in Fig.~\ref{fig:quality_bench_results}(a). As we increase the cost budget, the accuracy increases for both optimal and random parameters. Optimal parameters consistently outperform the random choice by $2\times$ to $4\times$.

\subsection{Evaluate with QPU experiments}
To further validate the method, we perform QPU experiments. Unlike numerical simulations, the ground truth of QPU quality is unknown. We perform a full RB experiment to characterize the Clifford gate fidelity of an edge on QPUs, and use the RB result as the ground truth quality, $\lambda_\text{QPU}$. We emphasize that this is a proxy rather than a true ground truth: the RB estimate carries its own statistical uncertainty and inherits the assumptions of the RB noise model. This is nevertheless the operationally relevant comparison, since RB is precisely the Tier~1 procedure the screen is designed to gate, and detecting deviations from the RB-characterized state is the screen's task in deployment. The RB experiments are done with 14 different Clifford sequence lengths, ranging from 5 to 130 Clifford gates, with 10 distinct random sequences for each length. The full RB experiment therefore consists of 140 circuits.

We compare 14 different screening designs, each consisting of a subset of the QPU data from the full RB experiment. Each screening design uses the 10 circuits with the same Clifford sequence length. It is assumed that the drifts in QPU quality follow a normal distribution centered at the spec quality, $\lambda_\text{spec}$. That is, $\lambda_\text{QPU} \sim \mathcal{N}(\lambda_\text{spec},\sigma_\text{drift}^2)$, and $\sigma_\text{drift}=0.01$ for the evaluation purpose. We apply the hypothesis testing formalism from Eq.~\eqref{eqn:t-test quality verification} and decision rule from Eq.~\eqref{eqn:t_statistic_decision_rule} with critical value $t_\alpha=2$ to compute the TPR and true negative rate (TNR, the ratio of true negatives to actual negatives) of each design of the screening probes. The average of TPR and TNR, balanced accuracy, is reported in Fig.~\ref{fig:quality_bench_results}(b) with three different QPUs, Rigetti Ankaa-3, IQM Garnet and IQM Emerald. The optimal design predicted with Section~\ref{sec:quality verification, optimal exp design} is 95 effective gates with $f_0=0.99$, which corresponds to Clifford sequence lengths $N_g$ of 61, 51 and 36 for the three devices, respectively. We fit the model in Eq.~\eqref{eqn:quality verification, differential t value} to the QPU experiment data. The maximum of the fitted curves, designs that best differentiate QPU and spec quality are with 95, 116 and 88 effective gates, largely agreeing with the predictions. The results from the Ankaa-3 device match perfectly with the prediction. Results from Garnet and Emerald show 22\% and 7\% off from the predictions, but still within 1 standard deviation from the predicted optimal gate count. The deviation may come from larger deviations between fidelity spec and the actual fidelity at the time of the experiment.

\subsection{Comparison with OED-optimized RB}
\label{sec:rb_vs_monitoring}

The preceding results demonstrate that our screening probe outperforms the ones with non-optimal parameters. A natural follow-up question, raised by prior work on optimal experiment design for RB~\cite{kwiatkowski2023,harper2019,itoko2021sampling}, is how the screening-probe sensor compares against an RB experiment whose own design has been optimized. To address this, we perform numerical simulation to compare the true positive rate (TPR) of two Tier~2 sensors under identical drift conditions: (1)~an RB-based sensor whose sequence-length allocation follows the OED strategy of Itoko and Raymond~\cite{itoko2021sampling}, and (2)~the screening-probe sensor proposed in this work.

The gate fidelity $\lambda_\text{QPU}$ drifts from its calibrated value $\lambda_\text{spec}$ according to $\lambda_\text{QPU} \sim \mathcal{N}(\lambda_\text{spec}, \sigma^2)$, where $\sigma$ is the drift standard deviation. We set $\lambda_\text{spec} = 0.97$ and sweep $\sigma \in [10^{-4},\, 10^{-2}]$.

\begin{itemize}
    \item \textbf{Baseline (OED-optimized RB):} Directly estimates $\lambda_\text{QPU}$ via randomized benchmarking with sequence lengths chosen according to~\cite{itoko2021sampling}. An alarm is raised when a $t$-test rejects the null hypothesis that the estimated $\lambda_\text{QPU}$ equals $\lambda_\text{spec}$. OED is done with fixed shots of 100 and $M_\text{max}=15$ (see Ref.~\cite{itoko2021sampling} for definition).
    \item \textbf{Proposed approach (OED-optimized screening probe):} Executes 68 probe circuits with 100 shots, each with optimal Clifford sequence length $m = 31$ ($=95$ effective gates with $f_0=0.99$), and observes the success probability, $S$. An alarm is raised when a $t$-test rejects the null hypothesis that $S$ equals the success probability predicted by $\lambda_\text{spec}$ via Eq.~\eqref{eqn:noise_model}.
\end{itemize}

\noindent The two approaches are compared at equal resource cost of running 68 distinct circuits each with 100 shots. 

Fig.~\ref{fig:quality_bench_results}(c) shows the TPR of both approaches as a function of drift rate $\sigma$. The proposed approach achieves substantially higher TPR across the entire range of $\sigma$. At $\sigma \approx 10^{-3}$, the proposed approach detects ${\sim}73\%$ of drift events while the baseline detects only ${\sim}5\%$. At $\sigma \approx 5 \times 10^{-3}$, the gap narrows but persists ($94\%$ vs.\ $69\%$, ${\sim}1.4\times$).

The key distinction is not in how the RB experiment is designed internally, but in \emph{what} is being measured. Prior OED-for-RB methods~\cite{kwiatkowski2023,harper2019,itoko2021sampling} optimize the allocation of sequence lengths and samples to minimize the variance of the fidelity \emph{estimator} $\hat{\lambda}$. This is the right objective when the goal is parameter estimation. However, for drift \emph{detection}, the goal is not to estimate $\lambda$ precisely but to determine whether $\lambda$ has changed. The screening-probe sensor exploits the power-law amplification $\lambda \mapsto \lambda^{m}$: near the nominal value $\mu$, a drift $\Delta\lambda$ produces a signal $\Delta S \approx m\,\mu^{m-1}\,\Delta\lambda$ in the success probability, which grows linearly with $m$. This amplification, combined with the optimal circuit-size selection from Section~\ref{sec:optimal_probe}, yields a fundamentally more sensitive detector than full RB characterization which first estimates $\lambda$ and then thresholds the estimate. At equal resource cost, this translates to a ${\sim}4.4\times$ higher detection rate at $\sigma \approx 1.5 \times 10^{-3}$ ($82\%$ vs.\ $18\%$), or equivalently, the same detection capability with proportionally fewer circuits and less QPU time.

\section{Application 2: Compiler benchmarking}
\label{sec:compiler_benchmarking}
Having demonstrated the strategy for quality verification of a single device parameter $\lambda$, we now apply the same optimization machinery to compiler benchmarking, where the monitored quantity is a \emph{relative} fidelity difference between two compilation strategies on a given device.

While a single compiler comparison might appear to be a one-time evaluation, several factors make it a recurring screening problem. First, noise-adaptive compilation strategies that exploit current calibration data can shift in relative ranking as device parameters evolve~\cite{kurniawan2024calibration,huo2025revisiting,xu2025qsteed}. Second, compiler software is updated regularly, and each update may alter the quality of compiled circuits. Third, large-scale applications may consist of a stream of structurally diverse quantum programs, and a single compilation strategy may not be optimal for all circuit types. These factors make the two-tier framing natural: a cheap screening probe triggers an expensive Tier~1 response only when a significant change in relative compiler performance is detected. In the compiler setting, the Tier~1 response goes beyond simply re-running a benchmark suite: it typically involves an investigation of compiler performance, including root-cause analysis, profiling of compilation passes, and potentially compiler improvements by the developers.

The two-tier architecture (Section~\ref{sec:two_tier}) offers a practical alternative. A small \emph{screening subset} of circuits, chosen to maximize sensitivity to compiler differences, serves as the Tier~2 screen. Only when this screen detects a statistically significant fidelity gap is the full compiler investigation triggered. Because the screening subset need not be representative of all workload types, only maximally discriminating, it can be much smaller than the full suite while still providing a reliable trigger.

\subsection{Hypothesis test}
\label{sec:compiler_hyp_test}

Given a target QPU and a class of workload circuits, two compilers $A$ and $B$ each produce compiled circuits whose execution yields output distributions.
The compiler-benchmarking question is: \emph{do compilers $A$ and $B$ produce circuits of different execution quality on the target device?}
A naive comparison based on a small number of compiled circuits is statistically unreliable: shot noise, device drift during the experiment, and circuit-to-circuit variance can all mask or mimic genuine compiler differences.
Following the template of Section~\ref{sec:general_hypothesis}, we formalize the problem as a two-sided hypothesis test.

Because both compilers act on the same source circuits and target the same device, a paired design is natural: for each source circuit $i$, we obtain a fidelity pair $(F_{A,i},\, F_{B,i})$ and form the paired difference $D_i = F_{A,i} - F_{B,i}$, where $A$ and $B$ denote the two compilers.
The hypotheses then reduce to a test on the mean of $D$: $H_0 : \mathbb{E}[D] = 0$ against $H_a : \mathbb{E}[D] \neq 0$.

While a single paired difference cannot isolate the individual compilation choices (routing, gate cancellation, scheduling) responsible for a fidelity gap, rejection of $H_0$ on the screening subset indicates that the two compilation strategies likely differ in execution quality on the full workload as well.
The test therefore serves as a first gate: only when $H_0$ is rejected is a full compiler investigation warranted.

Note that, rather than gate count, we use Hellinger fidelity between the measured and ideal distribution as the comparison metric. Typical compiler benchmarks rely on gate count~\cite{weiden2022wide, nation2025benchmarking}.
However, gate count is only a structural proxy and does not always reliably predict execution quality on real hardware.
Empirical studies show that established figures of merit can have weaker-than-expected correlation with measured execution quality~\cite{hopf2025improving}, and that logically equivalent circuits with similar structural footprints can yield up to order-of-magnitude differences in fidelity~\cite{dangwal2024compass,wu2021hierarchical,quetschlich2023predicting}.
Computing the ideal output distribution generally does not scale to large circuits, but mirror-circuit techniques and related scalable constructions can circumvent this limitation by designing circuits whose ideal output is known by construction~\cite{proctor2022scalable,mayer2021theory,erhard2019characterizing,blumekohout2020volumetric,proctor2022measuring}.

\subsection{Optimal experiment design}
\label{sec:compiler_oed}

Unlike quality verification, where the probe circuits are synthetic RB sequences designed from scratch (Section~\ref{sec:app_quality}), compiler benchmarking uses application circuits drawn from a benchmark suite. The design question is therefore not how to construct the probe circuit, but which circuits from the suite to include in the screening subset and, in particular, at what circuit size the fidelity difference between compilers is most detectable.

\subsubsection{Fidelity difference model}

Suppose an uncompiled circuit has $N$ effective gates (Section~\ref{sec:universality}).
After compilation by compilers $A$ and $B$, the effective gate counts become $N - \Delta N_A$ and $N - \Delta N_B$, respectively, with expected fidelities $F_A = f_0^{N - \Delta N_A}$ and $F_B = f_0^{N - \Delta N_B}$.
The fidelity difference is governed by the compiled gate-count difference $\Delta N_d = \Delta N_A - \Delta N_B$: expanding to first order in $\Delta N_d$ gives $\Delta F = F_A - F_B \approx (-\log f_0)\,\Delta N_d\, f_0^{N}$.
Under the assumption that optimizations operate locally, each compiler removes a number of gates proportional to $N$, so $\Delta N_d = \kappa N$ with $\kappa$ the difference in per-gate reduction between the two compilers. The fidelity difference then scales as
\begin{equation}
    \label{eqn:delta_F}
    \Delta F \approx \alpha\,N\,f_0^{N},
\end{equation}
where $\alpha = \kappa(-\log f_0)$ is a compiler-pair-dependent constant.
This expression reflects the same competing trends identified in Section~\ref{sec:optimal_probe}: the difference grows linearly with circuit size $N$ but decays exponentially through the noise factor $f_0^{N}$, yielding a maximum at a finite circuit size.

\subsubsection{Statistical formulation}

To assess statistical significance, we cast the compiler comparison as a paired $t$-test.
Given $K$ benchmark circuits compiled by both compilers and executed on the QPU, we compute the paired fidelity differences $\{\Delta F_k\}_{k=1}^{K}$ and form the $t$-statistic $t = \overline{\Delta F}/\sigma_{\Delta F}$.
For algorithm-level circuits whose ideal output distributions spread probability mass over many bitstrings, the standard error depends primarily on the total number of shots rather than circuit structure, so we approximate $\sigma_{\Delta F} \approx \sigma_0 / \sqrt{K}$.
The expected $t$-value under $H_a$ is then
\begin{equation}
    \label{eqn:t_compiler_expanded}
    t \approx \frac{\alpha\,N\,f_0^{N}}{\sigma_0 / \sqrt{K}}.
\end{equation}
Because $\sigma_0$ is approximately constant with respect to $N$, maximizing $|t|$ over $N$ reduces to maximizing $\Delta F(N) = \alpha\,N\,f_0^{N}$, the task-specific objective $\Phi$ in the general constrained optimization of Eq.~\eqref{eqn:general_optimization}.

\subsubsection{Optimal circuit size}
\label{sec:compiler_optimal_circuit_size}

The optimal circuit size that maximizes  Eq.~\eqref{eqn:t_compiler_expanded} is
\begin{equation}
    \label{eqn:optimal_circuit_size}
    N_m = \frac{1}{-\log f_0},
\end{equation}
identical in form to the quality-verification optimum of Eq.~\eqref{eqn:optimal_M_general}, demonstrating the generality of the strategy. The Fisher information analysis of the underlying broad-distribution model is given in Appendix~\ref{app:fisher_broad}.
For $f_0 = 0.99$, this yields $N_m \approx 100$ effective gates, independent of the compiler-dependent constant $\alpha$.

The useful range can be estimated from the standard deviation of Eq.~\eqref{eqn:delta_F}, $\sigma_N = \sqrt{2 f_0}/(1 - f_0)$. For $f_0 = 0.99$, $\sigma_N \approx 140$, so the window $N_m \pm \sigma_N$, two standard deviations wide, spans roughly 1 to 240 effective gates.
Circuits with small effective gate count are compiled near-perfectly by any compiler; circuits above roughly 240 effective gates degrade into noise regardless of compilation quality.
This range defines the screening subset: circuits whose effective gate counts fall within the optimal window are the most informative probes for the Tier~2 screen. The number of circuits in the subset and the shots per circuit can be determined via the constrained optimization of Eq.~\eqref{eqn:general_optimization} for a given resource budget.

\subsection{Evaluation with numerical and QPU experiments}
\label{sec:compiler_results}

We evaluate the optimal circuit-size prediction of Section~\ref{sec:compiler_optimal_circuit_size} through noise simulations and QPU experiments. The central question is whether the fidelity difference between compilers is indeed maximized at the predicted optimal effective gate count, and whether circuits in this range provide the most sensitivity for distinguishing compilation strategies.

\paragraph{Experimental setup}
\label{sec:compiler_setup}

Three compilers were evaluated: Tket (custom routing, mapping, and optimization passes), Qiskit (default layout and routing with level-3 optimization), and UCC (default options). All compilers make routing and mapping decisions that depend on the characterization of each target device. The benchmark suite includes standard algorithm families (Bernstein--Vazirani, GHZ, phase estimation, QAOA, quantum volume, and TFIM) with qubit counts from 2 to 20, spanning a wide range of effective gate counts.

\begin{figure}[h]
\centering
\includegraphics[width=0.8\linewidth]{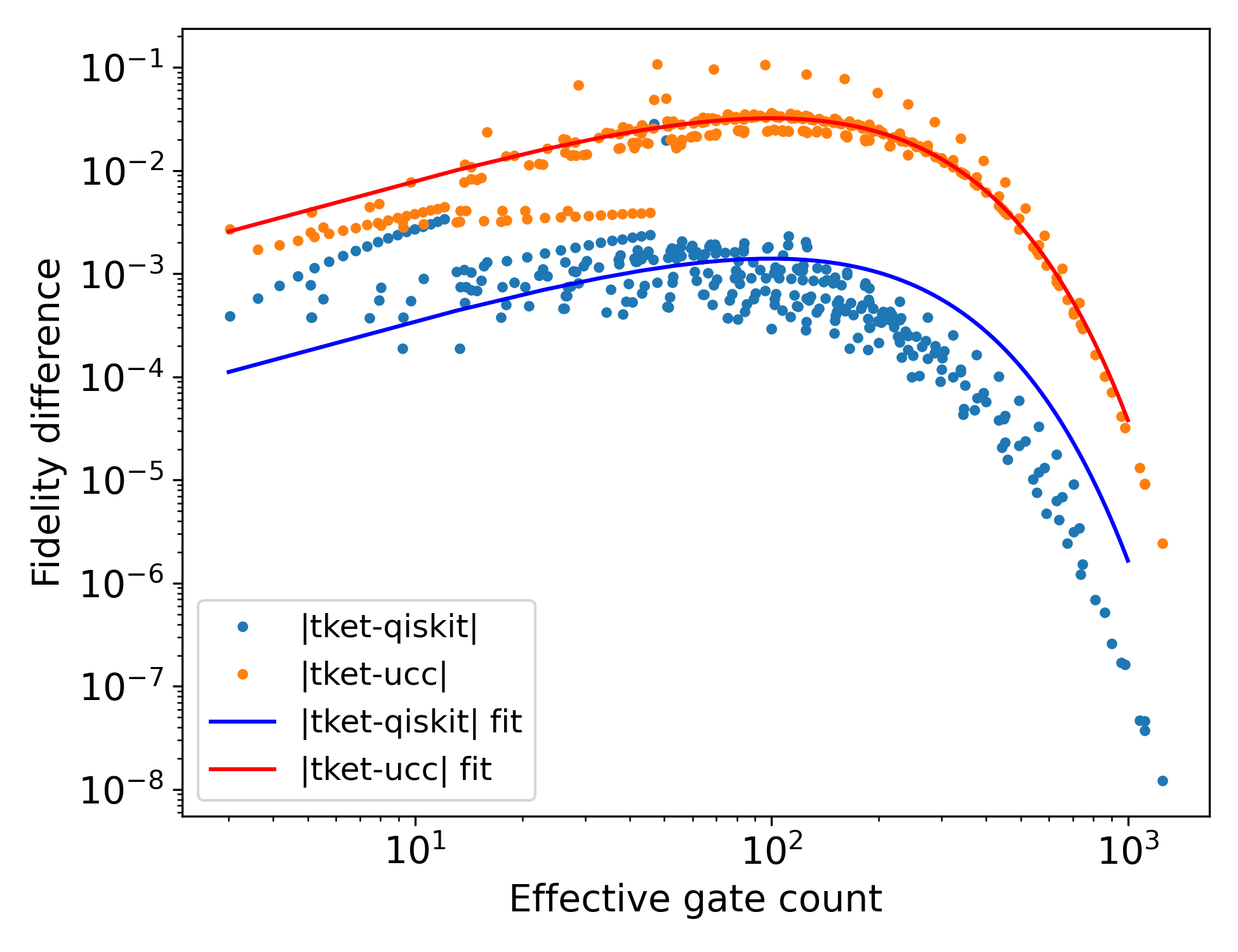}
\caption{Absolute fidelity difference between circuits compiled by different compilers as a function of effective gate count. Each data point is the numerically simulated fidelity difference of a single circuit. The solid line is the fit of Eq.~\eqref{eqn:t_compiler_expanded}.}
\label{fig:compiler_benchmarking_fidelity_diff_expected}
\end{figure}

\paragraph{Numerical simulations}
Fig.~\ref{fig:compiler_benchmarking_fidelity_diff_expected} shows the numerically simulated fidelity difference between circuits compiled by Qiskit and Tket, and between circuits compiled by UCC and Tket, targeting IonQ Aria-1. For both compiler pairs, the largest difference in circuit fidelities occurs around 100 effective gates, confirming the prediction of Section~\ref{sec:compiler_optimal_circuit_size}. Circuits with fewer than ${\sim}30$ effective gates show negligible fidelity differences (both compilers produce near-ideal results), while circuits above ${\sim}240$ effective gates are dominated by noise, diminishing compiler differences. The solid curves show the fit of Eq.~\eqref{eqn:t_compiler_expanded}, which captures the rise-and-fall behavior well.

\begin{figure}[h]
\centering
\includegraphics[width=0.8\linewidth]{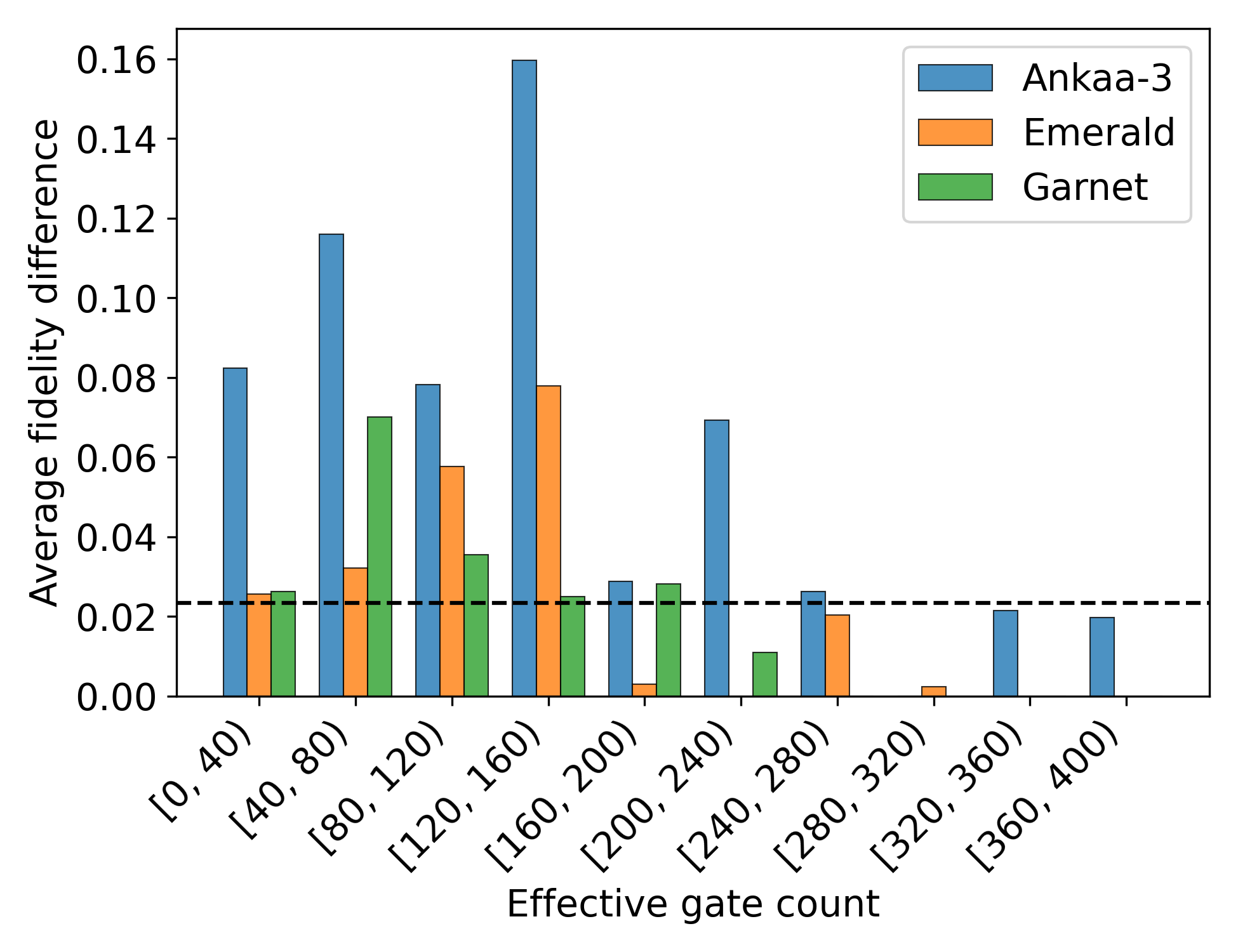}
\caption{Absolute Hellinger fidelity difference between Tket-compiled and Qiskit-compiled circuits from QPU experiments. The dashed line indicates twice the standard error estimated by noise simulation. The largest differences concentrate in the 40--160 effective gate range, validating the optimal circuit-size model.}
\label{fig:compiler_benchmarking_fidelity_diff_qpu}
\end{figure}

\paragraph{QPU experiments}
We validate the predictions on three superconducting devices: Rigetti Ankaa-3, IQM Garnet, and IQM Emerald. Fig.~\ref{fig:compiler_benchmarking_fidelity_diff_qpu} shows the absolute Hellinger fidelity difference between Tket-compiled and Qiskit-compiled circuits. The most discriminating gate-count ranges were 120--160 effective gates for Ankaa-3, 40--80 for Garnet, and 120--160 for Emerald. Across all devices, the largest fidelity differences concentrate in the 40--160 effective gate range, consistent with the theoretical prediction.

The numerical and QPU results jointly confirm that Eq.~\eqref{eqn:t_compiler_expanded} reliably identifies the circuit-size range where compiler differences are most detectable. For practitioners, this means that a screening subset drawn from this range will maximize the probability of detecting a genuine compiler regression, while circuits outside this range contribute little discriminating power and can be excluded from the screening step to save resources.

\section{Discussion}
\label{sec:discussion}

The strategy presented here applies to any benchmarking task that can be cast as a hypothesis test with a tunable probe circuit under an exponential fidelity model. The two applications presented in this paper both fall under this model, which is what enables the closed-form optimal circuit sizes and their convergence to the same expression as a function of design parameters. For benchmarking tasks governed by different noise models (e.g., non-exponential decay or correlated errors), the two-tier architecture and the sensitivity-maximization principle still apply, but the optimal probe design would need to be re-derived for the specific model. 

In deployment, the false-negative risk of the Tier~2 screen (Section~\ref{sec:two_tier}) is managed operationally. The operator chooses the smallest drift magnitude considered actionable, for example the fidelity degradation that would violate an application's error budget, and a target detection probability for it. The constrained optimization of Eq.~\eqref{eqn:general_optimization} then determines whether the available budget suffices or must be raised. Two properties of the architecture further limit the risk. Because screening repeats at every interval and drift accumulates over time, a drift missed at one interval presents a larger signal at the next, so a false negative manifests as detection latency rather than a permanent miss. In addition, the screen can be complemented by an occasional unconditional Tier~1 run as a safety margin, capping the worst-case time a slowly drifting device operates out of specification, mirroring common practice in risk-based testing~\cite{felderer2014taxonomy}.

Several limitations should be noted. The two-tier model assumes instantaneous and perfect sensing and calibration; in practice, the sensing step has finite duration and measurement uncertainty, and calibration may not perfectly reset the system. The finite probe duration is partially captured by the framework itself, since the probe cost and its measurement uncertainty enter the constrained optimization of Eq.~\eqref{eqn:general_optimization}, but imperfect calibration resets would require a residual-error term in the drift model. The exponential fidelity model assumes time-invariant, gate-independent, depolarizing noise, assumptions that are standard in RB analysis but approximate on real hardware. Correlated errors, non-Markovian dynamics, and spatially varying noise would require extensions to the EGC reduction. On the experimental side, the QPU validation uses full RB as a proxy ground truth since the physical fidelity is not directly observable; the reported accuracies therefore quantify consistency with the Tier~1 procedure rather than with an absolute reference (Section~\ref{sec:app_quality}). The framework does not address dependency propagation across a calibration graph, as handled by Optimus~\cite{kelly2018physical}; however, the optimally designed probe derived here could serve as a drop-in replacement for the heuristic \texttt{check\_data} step in such architectures. Finally, the Fisher information analysis identifies the optimal design point but does not account for model misspecification; robustness to deviations from the assumed noise model is an important direction for future work. We note, however, that model misspecification degrades the screen gracefully: a suboptimal probe size reduces detection power but does not invalidate the hypothesis test itself, and the false-negative safeguards discussed above remain in force.

\section{Conclusion}

We presented a resource-optimal screening strategy for quantum system benchmarking that combines a two-tier, risk-based testing architecture with optimal experiment design. The strategy uses a lightweight screening probe to detect statistically significant changes in system performance, triggering expensive full characterization only when needed. Under an exponential fidelity model, we derived closed-form optimal circuit sizes that maximize detection sensitivity, and introduced the effective gate count to make the analysis hardware-independent.

We demonstrated the strategy on two applications. For quality verification, the optimally designed screening probe achieves a $4\times$ higher detection rate than OED-optimized RB at equal resource cost. For compiler benchmarking, the derived optimal circuit size correctly predicts the most discriminating circuit-size range on hardware. Hardware and numerical experiments combined, four QPUs (Rigetti Ankaa-3, IQM Garnet and Emerald, IonQ Aria-1) are targeted to validate the theoretical predictions.

Looking ahead, a natural next application is error-mitigation validation: detecting when a learned noise model has become stale and needs re-characterization. Similarly, in QEC, decoders that exploit device-specific error models achieve lower logical error rates than generic decoders~\cite{chen2022calibrated,battistel2023real}; monitoring whether the error model has drifted sufficiently to degrade decoding performance could trigger targeted re-characterization, which may be important for fault-tolerant quantum computing applications. In both cases, the two-tier architecture and sensitivity-maximization machinery carry over directly; only the observable $Y$ and quality parameter $\theta$ change. As quantum processors scale and calibration overhead grows, we expect that resource-optimal screening will become increasingly important for maintaining system performance without sacrificing computational throughput.

\section*{Acknowledgments}
The authors thank Mao Lin, P{\'e}ter K{\'o}m{\'a}r, and Eric M. Kessler for valuable discussions. The authors used Amazon Quick for grammar and spelling correction and for rewording sentences to improve readability.

\input{appendix_fisher_v1_short}

\bibliographystyle{IEEEtran}
\bibliography{reference}

\end{document}

%% file: appendix_fisher_v1_short.tex
\appendices

\section{Fisher Information Derivations}
\label{app:fisher}

We summarize the modeling and the key Fisher-information results used in the main text. The derivations follow standard optimal experiment design techniques~\cite{Pukelsheim2006,VanTreesBellTian2013}, and we state the results needed to reproduce the closed-form optima.
A benchmarking experiment samples outcomes $x \sim p(x \mid \theta_K, \theta_U)$, where $\theta_K$ are controllable design variables (e.g., circuit size $M$) and $\theta_U$ is the unknown quality parameter (e.g., the gate fidelity $p$). Optimal experiment design chooses $\theta_K$ to maximize the Fisher information
\begin{equation}
    \mathcal{I}(\theta_U;\theta_K) = \mathbb{E}\!\left[\bigl(\partial_{\theta_U}\!\log p(x\mid\theta_K,\theta_U)\bigr)^{2}\right],
\end{equation}
which, through the Cram\'{e}r--Rao bound $\mathrm{Var}(\hat{\theta}_U)\!\ge\!\mathcal{I}^{-1}$, lower-bounds the variance of any unbiased estimator independently of the estimator used~\cite{Pukelsheim2006,VanTreesBellTian2013}. Under the canonical global-depolarizing, time-invariant noise model used in the main text, each measurement outcome is distributed as a mixture parameterized by $t(p) = (1-c)p^{M} + c$, so the circuit size $M$ is the single design knob that controls statistical sensitivity. We analyze two limiting cases that bracket the benchmarking tasks of interest: a peaked distribution concentrated on the ideal outcome (quality verification) and a broad distribution spread across many outcomes (compiler benchmarking).

\subsection{Peaked distribution (quality verification)}
\label{app:fisher_peak}

For a Bernoulli outcome with success probability $q(p) = (1-c)p^{M} + c$ and $N$ i.i.d.\ trials, the Fisher information in $p$ is
\begin{equation}
    \mathcal{I}_N(p;M,c)
    = \frac{N(1-c)\,M^{2}\,p^{2M-2}}{\bigl(c+(1-c)p^{M}\bigr)\bigl(1-p^{M}\bigr)}.
\end{equation}
Maximizing over $M$ yields the closed-form optimal circuit size
\begin{equation}
    \label{eqn:Mstar_app}
    M^{\star}(p,c) = \frac{x^{\star}(c)}{-\ln p},
\end{equation}
where $x^{\star}(c)$ solves a scalar first-order condition. For $c = 0$, $x^{\star} = 2 + W_0(-2/e^{2}) \approx 1.594$ with $W_0$ the principal Lambert-$W$ branch. The readout floor $c$ compresses the dynamic range of the observable and shifts the optimum: with $c = 0.25$ (the depolarization floor used in the main text), $x^{\star} \approx 0.96$, so for $f_0 = 0.99$ the optimum is $M^{\star} \approx 95$ effective gates. 

\subsection{Broad distribution (compiler benchmarking)}
\label{app:fisher_broad}

For a circuit with $m$ outcomes whose probabilities follow $q_i(p) = t(p)\,r_i + (1-t(p))\,u_i$, the multinomial Fisher information in the full count vector $C \sim \mathrm{Multinomial}(N,q)$ is
\begin{equation}
    \mathcal{I}_C(p)
    = N\,(t'(p))^{2}\sum_{i=1}^{m}\frac{(r_i-u_i)^{2}}{q_i(p)}.
\end{equation}
In practice only the Hellinger fidelity $Y = F(C)$ to the ideal distribution is recorded, and the data-processing inequality gives $\mathcal{I}_Y(p)\le \mathcal{I}_C(p)$. No closed form exists for $\mathcal{I}_Y(p)$ because $F$ is a nonlinear many-to-one map, but a large-$N$ delta-method approximation yields
\begin{equation}
    \mathcal{I}_Y(p) \approx \frac{N\,(\mu'(p))^{2}}{V(p)},
\end{equation}
with $\mu(p) = \mathbb{E}[Y]$ and $V(p)$ the delta-method variance. Numerically, $\mathcal{I}_Y(p)$ peaks at the same structural optimum $M^{\star}\propto 1/(-\ln p)$ as the peaked case, differing only in the prefactor, which is why both applications in the main text share the closed-form shape of Eq.~\eqref{eqn:Mstar_app}.

\section{First-Order Expansion of the Probe $t$-Value}
\label{app:t_value_expansion}

Under the small-drift assumption of Section~\ref{sec:quality verification, optimal exp design}, the $t$-value of the quality-verification probe expands to first order as
\begin{align} \label{eqn:t_value}
t & = \frac{\bar{Y}-S_\text{spec}}{\sigma} \notag 
 = \frac{
    \frac{\partial S}{\partial \lambda} \Delta\lambda 
    + \frac{\partial S}{\partial a} \Delta a
}{\sigma} 
\Big|_{(\lambda_\text{spec}, a_\text{spec}, N_g)} \notag \\
& = (1-c)\frac{
    N_g a \lambda^{N_g-1}\Delta\lambda + \lambda^{N_g}\Delta a
}{\sigma}
\Big|_{(\lambda_\text{spec}, a_\text{spec})} .
\end{align}
Today, most quantum hardware has $\lambda \simeq 0.99$ and $a \simeq 0.99$. From Eq.~\eqref{eqn:t_value}, the ratio between the contribution of $\Delta a$ and $\Delta\lambda$ is $\lambda^{N_g} / (N_g a \lambda^{N_g-1}) \simeq 1/N_g \ll 1$, where the approximation holds because $\lambda$ and $a$ are of the same order of magnitude. The $\Delta a$ term can therefore be neglected, yielding Eq.~\eqref{eqn:t_value_2} of the main text.